\documentclass{elsart}
\usepackage{graphicx}
\begin{document}
\begin{frontmatter}
\title{
 Best mean-field for condensates }
\author{L.S.\ Cederbaum and A.I.\ Streltsov}
\address{Theoretische Chemie, Universit\"at Heidelberg, D-69120 Heidelberg, Germany}
\date{\today}

\begin{abstract}
The Gross-Pitaevskii equation assumes that all (identical) bosons of a condensate reside in a single
one-particle function. Here, we raise the question whether it always provides the best mean-field ansatz
for condensates, leading to the lowest mean-field ground state energy.
To this end, we derive a mean-field approach allowing for bosons to reside in several different one-particle functions.
The number of bosons in each of these functions is a variational parameter minimizing the energy.
The energy and one-particle functions at these optimal numbers can be determined directly.
A numerical example is presented demonstrating that the mean-field energy of trapped bosons can be below that provided
by the Gross-Pitaevskii equation. Implications are discussed.
\end{abstract}

\begin{keyword}
\PACS{03.75.Hh,03.65.Ge,03.75.Nt}
\end{keyword}
\end{frontmatter}

Non-linear Schr\"odinger equations are widely used in many areas of physics and chemistry.
In fact, the mean-field equation of any many-body system is a non-linear Schr\"odinger equation.
Usually, however, one speaks of {\it the} non-linear Schr\"odinger equation if this equation takes on a 
particular form. This equation is used to describe many phenomena like solitons \cite{1},
optical pulses in fibres \cite{2}, Bose-condensed photons \cite{3} and others.
Motivated by famous experiments \cite{4,5,6}, extensive use of this equation has been made in resent years
to describe dilute  gas Bose-Einstein condensates \cite{7}. In the context of such condensates this equation
is also referred to as Gross-Pitaevskii (GP) equation \cite{8,9}.
  
The non-linear Schr\"odinger equation can be derived as the mean-field  equation of a system of N identical
bosons which interact via a $\delta$-function potential $ W(\vec{r}_1-\vec{r}_2)=\lambda_0\,\delta(\vec{r}_1-\vec{r}_2) $,
where $\vec{r}_i$ is the position of the i-th particle. It is equivalent to the GP equation
if we identify the appearing non-linear parameter with the s-wave scattering length of the particles \cite{7}.
At zero temperature all bosons can occupy  the same spatial one-particle function (orbital) if their mutual
interaction is neglected. The mean-field description of the interacting system is obtained by assuming the ground state
wave function $\Psi$ to be a product of identical spatial orbitals $\varphi$:
$\Psi(\vec{r}_1,\vec{r}_2,\ldots,\vec{r}_N)= \varphi(\vec{r}_1)\varphi(\vec{r}_2)\cdots\varphi(\vec{r}_N)$.
The energy $E\,\equiv\,<\Psi|\hat{H}|\Psi>$ defined as the expectation value of the widely used \cite{7,13}
Hamiltonian $\hat{H}= h + W$, reads
\begin{equation}
E_{GP}=N\{\int\varphi^* h\,\varphi\,d\vec{r}+\frac{\lambda}{2}\int|\varphi|^4\,d\vec{r}\},\lambda=\lambda_0(N-1) 
\end{equation}
where $ h(\vec{r})=\hat{T}\,+\,\hat{V}(\vec{r}) $ is the unperturbed one-particle Hamiltonian consisting of the
kinetic operator $ \hat{T} $ and the external potential $\hat{V}(\vec{r})$. By minimizing this energy one readily 
obtains the GP equation for the optimal orbital $ \varphi $:
\begin{equation}
\{\:h(\vec{r})+\lambda|\varphi(\vec{r})|^2\}\,\varphi(\vec{r})=\,\mu_{GP}\,\varphi(\vec{r})  
\end{equation}
Solution of this eigenvalue equation determines this $\varphi$ and
$\mu_{GP}$ which is called the chemical potential \cite{7}.

The GP equation and energy are very appealing and have been successful in explaining experimental observations.
We raise the question, however, whether the GP approach always provides the best mean-field description of the system.
We shall demonstrate below that {\it other} mean-fields may provide a lower energy than (1) does, at least in some cases.
The general wave function of N non-interacting bosons is a product 
$\varphi_1(\vec{r}_1)\varphi_2(\vec{r}_2)\cdots\varphi_N(\vec{r}_N)$ of orbitals which can all be different.
Since the bosons are identical, this product must, of course, be symmetrized and several bosons may occupy 
the same spatial orbital. We may put $n_1$  bosons in orbital $\varphi_1$,  $n_2$ in orbital $\varphi_2$ and so on.
By definition the most general mean-field energy is $E=<\Psi|\hat{H}|\Psi>$ where $\Psi$ is the above 
symmetrized product. For transparency of presentation we confine ourselves in the following to two orbitals
$\varphi_1$ and $\varphi_2$ with particle occupations $n_1$ and $n_2$, respectively, where $n_1+n_2=N$.
The extension to more orbitals is trivial.  The energy now reads
\begin{eqnarray}
 E = n_1 h_{11} + n_2 h_{22} + \lambda_0 \frac{n_1(n_1-1)}{2}\int|\varphi_1|^4 d\vec{r}+ 
 \nonumber  \\  \lambda_0 \frac{n_2(n_2-1)}{2}\int|\varphi_2|^4 d\vec{r}+ 
2 \lambda_0 n_1 n_2 \int|\varphi_1|^2 |\varphi_2|^2 d\vec{r} 
\end{eqnarray}
where $h_{ii}=\int\,\varphi_i^*\,h\,\varphi_i\,d\vec{r}$ is the usual one-particle energy. Minimizing 
this energy with respect to $ \varphi_1 $ and $ \varphi_2 $  under the constraints that they are orthogonal, 
$ <\varphi_1|\varphi_2>=0$, and normalized to unity, $ <\varphi_i|\varphi_i>=1$, we obtain the following equations
\begin{eqnarray}
\{\:h(\vec{r})+\lambda_0 (n_1-1)|\varphi_1(\vec{r})|^2+2 \lambda_0 n_2|\varphi_2(\vec{r})|^2 \}\,\varphi_1(\vec{r}) =  \nonumber & & \\ 
= \mu_{11}\,\varphi_1(\vec{r})+\mu_{12}\,\varphi_2(\vec{r}) \nonumber & & \\
\{\:h(\vec{r})+\lambda_0 (n_2-1)|\varphi_2(\vec{r})|^2+2 \lambda_0 n_1|\varphi_1(\vec{r})|^2 \}\,\varphi_2(\vec{r}) = \nonumber & & \\
= \mu_{22}\,\varphi_2(\vec{r})+\mu_{21}\,\varphi_1(\vec{r}) & &  
\end{eqnarray}
for the optimal orbitals. 
These equations should not be confused with those for condensates made of two types of bosons.
The $\mu_{ij}$ are the Lagrange parameters due to the above mentioned constraints.
Of course, the energy E and the orbitals $\varphi_1$,$\varphi_2$ depend on the particle occupation $n_1$.
Consequently, we may treat $n_1$ as a variational parameter and search for its optimal value which minimizes the energy.
We shall return to this central point below.

Here, we briefly make contact with the problem of fragmentation often discussed in the literature, 
see refs. \cite{a1,a2} and references therein.
A condensate is fragmented if its reduced one-body density matrix has two or more macroscopic eigenvalues.
Until now no fragmentation has been found for confined condensates, i.e., for condensates in an external potential.
Moreover, it has been shown that fragmentation cannot take place in a harmonic external potentials \cite{b}.
Our ansatz discussed above in eqs.(3,4) can describe fragmentation on the mean-field level. If the particle occupations
$n_1$ and $n_2$ minimizing the energy are both macroscopic for large N, fragmentation takes place.

Let us analyze eqs.(3,4). Using eq.(1) it follows that the energy needed to remove a particle without
changing the orbital $\varphi$ is, within the GP mean-field, given by $E_{GP}(N)-E_{GP}(N-1)=\mu_{GP}$.
Analogously, we can compute the energy needed to remove a particle from the orbital  $\varphi_1$ and that from
the orbital  $\varphi_2$ in the framework of the present mean-field. Using eqs.(3,4) and remembering that $\varphi_1$
and $\varphi_2$  are orthogonal, we readily find: $\mu_{11}=E(n_1,n_2)-E(n_1-1,n_2)$ and 
$\mu_{22}=E(n_1,n_2)-E(n_1,n_2-1)$. In other words, $\mu_{11}$ and $\mu_{22}$ can be viewed as 
{\it chemical potentials} of the $\{ \varphi_1 \}$- and  $\{ \varphi_2 \}$-particle manifolds, respectively.
 
Another interesting property of (3,4) is that, unless $n_1=n_2=N/2$, these equations are not invariant to
rotations (unitary transformations) of the pair $(\varphi_1,\varphi_2)$. This is due to the fact that $\varphi_1$ and
 $\varphi_2$ appear a different number of times in the wave function in contrast to the situation for
identical fermions where the mean-field equations (Hartree-Fock) are indeed invariant to rotations \cite{10}. 
As an important consequence the off-diagonal Lagrange parameters $\mu_{12}$ and $\mu_{21}$ can not, in general, be 
removed from eq.(4) by an unitary transformation. Finally we mention that $n_1\mu_{12}=n_2\mu_{21}$.

The factor $2\lambda_0 n_1 n_2 $ appearing in the energy (3) and the corresponding term in (4) deserve attention.
If the bosons residing in $\varphi_1$ would be distinguishable from those living in $\varphi_2$,
this factor will reduce to $\lambda_0 n_1 n_2 $. The remaining factor $\lambda_0 n_1 n_2 $ appears because of
the {\it exchange} interaction between the two manifolds of identical bosons. Concentrating on many
bosons, i.e. $n_1,n_2 \gg 1$, equations (3,4) become more transparent.
Then we may introduce the orbitals  $\phi_1=(n_1/N)^{1/2}\,\ \varphi_1$ and $\phi_2=(n_2/N)^{1/2}\,\ \varphi_2$
which are, of course orthogonal, but now normalized to $n_1/N$ and $n_2/N$, i.e., to the fraction of particles 
residing  in $\varphi_1$ and $\varphi_2$, respectively.
The energy (3) can now be expressed as

\begin{eqnarray}
 E  &=& E_{cl} + E_{ex} \nonumber  \\
 E_{cl} &=& \bar{h}_{11}+\bar{h}_{22}+\frac{\lambda}{2}\int\rho^2(\vec{r})d\vec{r}\\
 E_{ex} &=&\lambda \int  |\phi_1|^2 |\phi_2|^2 d\vec{r} \nonumber 
\end{eqnarray} 

where $\bar{h}_{ii}$ is defined as $h_{ii}$ in (3), but with the orbitals $\phi_i$ and \\
$\rho=\left[n_1|\varphi_1|^2+n_2|\varphi_2|^2\right]/N=|\phi_1|^2+|\phi_2|^2$ \\
is the density per particle.  It is easy to see that the 'classical' energy $E_{cl}$  
is invariant to unitary transformation of the pair 
$(\phi_1,\phi_2)$ while the exchange energy $E_{ex}$ is not. It is the latter energy which makes the situation interesting.
Without the exchange energy also the equations for the optimal orbitals, eq.(4), would be invariant to
unitary transformations. 

In the following we demonstrate that the mean-field energy (3) can be superior to the GP energy (1).
Our example consists of an one-dimensional attractive condensate $(\lambda_0<0)$ in an external symmetric double-well 
potential. We choose as potential $(a\geq |x_0|)$  
\begin{equation}
V(x)=c+\frac{1}{2}\omega x^2-\omega a \left[x^2-x_0^2+a^2\right]^{1/2}  
\end{equation}
 where the bottoms of the well are at $x=\pm x_0$.
The parameter $a$ is chosen such that $V(\pm x_0)=0$. The corresponding kinetic energy reads 
$\hat{T}=-\frac{\omega}{2}\frac{\partial^2}{\partial x^2}$ implying that $x$ is a dimensionless
coordinate. Since the potential is symmetric, the ground state of $h$ is a gerade function and the first 
excited state is an ungerade function of $x$. We, therefore, consider in our mean-field $\varphi_1$ and $\varphi_2$ 
to be gerade and ungerade functions of $x$, respectively. As can be seen from eq.(4) one finds $\mu_{12}=\mu_{21}=0$
because $\varphi_1$ and $\varphi_2$ possess different spatial symmetries. The equations for 
the optimal orbitals have been evaluated numerically, by solving the resulting eigenvalue 
equations self-consistently using DVR \cite{11} and finite-difference \cite{12} methods.
Both methods have led to identical results.


Using the same numerical procedures we have also solved the GP equation (2) and determined $\varphi$.
Having found the optimal orbitals, we have computed the mean-field energies $E$ and $E_{GP}$.
Figure 1 shows the difference of the energies per particle $(E-E_{GP})/N$ as a function 
of the fractional occupation $n_1/N$ for several values of the coupling constant $\lambda=\lambda_0 N$
(note that $N\gg1$). At $n_1/N=1$ the energy E, of course, coincides with the GP energy $E_{GP}$ 
for all values of $\lambda$. For $\lambda=0.5$ the energy $E$ grows as $n_1/N$ is decreased.
This is also true for $\lambda=-0.5$, but the growth of $E$ is less pronounced. Until about $\lambda\approx-1.0$,
$E>E_{GP}$ implying that the GP mean field is the best mean field and no particle occupy $\varphi_2$. 
At $\lambda\leq-1.0$ the scenario changes as the energy $E$ drops below $E_{GP}$ 
as a function of the fractional occupation  $n_1/N$  for a given value of  $\lambda$. The optimal mean field
is obtained at the minimum of the curve $E=E(n_1/N)$. This minimum is about $n_1/N=3/4$ for $\lambda=-1.5$
and approaches  $n_1/N=1/2$ continuously as $\lambda$ is further decreased to $\lambda=-4.0$. 
The energy gain per particle due to the present mean field is substantial; for the same potential the
energy difference between the ground and first excited states of $h=T+V$ is $\sim0.18$ and the height of
the barrier separating the two wells is $\sim1.0$.

The results are better understood by inspecting fig.2 which shows the density per particle $\rho(x)$
for $\lambda=-2.5$  and $n_1/N=0.6$.
As can be seen in fig.1, this value of $n_1/N$ corresponds to the minimum  of the energy for $\lambda=-2.5$.
The GP density per particle $|\varphi(x)|^2$ is also shown in fig.2 for comparison. 
It is eye catching that much of $|\varphi(x)|^2$ has moved from the center of the potential $x\approx0.0$ to its wells as 
an impact of the present mean field. Furthermore, the two peaks of the density have narrowed. The density
$\rho(x)$  has substantially localized in comparison to the GP density. Since the condensate is attractive $(\lambda<0)$,
localization is well-known to lead to a lowering of the energy \cite{13}. 
What is surprising is that localization beyond
the symmetry preserved GP density can be achieved in the framework of a mean-field approach 
without breaking the symmetry and without the necessity to go beyond the mean-field world.
We remark that the system should have more than two bosons to achieve this localization: because
of the different spatial symmetries of $\varphi_1$ and $\varphi_2$, their product is not gerade as the ground state
is expected to be. If more bosons are present, an even number of them can reside in $\varphi_2$.
  

The energy $E$ has been discussed above as a function of $n_1$. The particle  occupation $n_1$ 
is a variational parameter and we may ask whether the minimum of $E=E(n_1)$ can be determined directly
without computing first the energy as a function of $n_1$ and subsequently searching for the minimum. 
The answer is positive. We define a new function
\begin{equation}
\tau(n_1,n_2)=E(n_1,n_2) + \nu(n_1+n_2-N) 
\end{equation} 
of $n_1$ and $n_2$ and the constraint $n_1+n_2=N$ is represented by the Lagrange parameter $\nu$.
At its extrema $\partial \tau/\partial n_1=\partial \tau/\partial n_2=0$, leading to two relations,
one of which fixes $\nu$ and other  gives rise to the equality 
$\partial E/\partial n_1=\partial E/\partial n_2$. Taking the derivative of $E$ in (3) with respect to $n_1$ and 
$n_2$ for $n_1,n_2\gg 1$ recovers the expressions for $\mu_{11}$ and $\mu_{22}$ determined from (4), i.e.,
$\mu_{ii}=\partial E/\partial n_i, i=1,2$. Consequently we find that the two quantities $\mu_{11}$ and $\mu_{22}$
are equal at the extremum of $E$:
\begin{equation}
\mu_{11}=\mu_{22}\equiv\mu
\end{equation} 
This is a relevant finding for understanding the concept of a condensate and, furthermore, 
allows us to transform the basic eqs.(4) at the optimal occupation $n_1$ as discussed in the following.

We now introduce the new orbitals
\begin{eqnarray}
 \psi_1(\vec{r} ) & = &(\frac{n_1}{N})^{1/2}\varphi_1(\vec{r})+(\frac{n_2}{N})^{1/2}\varphi_2(\vec{r}) \nonumber \\
 \psi_2(\vec{r} ) & = &(\frac{n_2}{N})^{1/2}\varphi_2(\vec{r})-(\frac{n_1}{N})^{1/2}\varphi_1(\vec{r})  
\end{eqnarray} 
and express (4) by them. 
Confining ourselves to the minimum of $E$ and using the basic relation (8), the resulting equations for
$\psi_1$ and $\psi_2$ simplify considerably:
\begin{eqnarray}
\{\:h(\vec{r})+\frac{3}{4}\lambda|\psi_1|^2+\frac{1}{4}\lambda|\psi_2|^2 \}\,\psi_1 & = &\,
(\mu+\bar{\mu})\,\psi_1 \nonumber \\
\{\:h(\vec{r})+\frac{3}{4}\lambda|\psi_2|^2+\frac{1}{4}\lambda|\psi_1|^2 \}\,\psi_2 & = &\,
(\mu-\bar{\mu})\,\psi_2 . 
\end{eqnarray}

The equations (4) for the orbitals $\varphi_1$ and $\varphi_2$ are, in general, not eigenvalue equations. The corresponding
equations for $\psi_1$ and $\psi_2$ at the optimum occupation $n_1$ are eigenvalue equations. 
The eigenvalues are $(\mu+\bar{\mu})$ and $(\mu-\bar{\mu})$, where $\mu$ is the condensate's chemical potential (see (8))
and  $\bar{\mu}\equiv(n_1/n_2)^{1/2}\mu_{12}=(n_2/n_1)^{1/2}\mu_{21}$.

At first sight it is intriguing to see that the new eqs.(10) do not exhibit the optimal $n_1$ and $n_2$. Nevertheless, these
occupations can be deduced from these equations. To this end it is relevant to notice that (9)
is not a unitary transformation. Indeed, $\psi_1$ and $\psi_2$ are normalized to unity, 
$<\psi_1|\psi_1>=<\psi_2|\psi_2>=1$, but they are {\it not} orthogonal. Their scalar product is
\begin{equation}
<\psi_1|\psi_2>=\frac{n_2-n_1}{N}
\end{equation} 
and determines the optimal occupations. One can solve the two eigenvalue equations (10)
self-consistently and compute the optimal $n_1$ and $n_2$ from the final result.
We have tested this procedure for several numerical examples including the above discussed double-well case.
One directly obtains the minima of the energy curves shown in fig.1.
A practical hint: numerical stability is enhanced if at each iteration step use is made of the relationship (9)
between $\psi_1$ and $\psi_2$.

Let us conclude. By expressing the wave functions as a symmetrized product of single-particle functions (orbitals),
equations for the mean-field energy and the optimal orbitals are derived.
Several bosons can reside in each of these orbitals  $\varphi_i$ and it is a basic property of the present approach
that these occupation numbers $n_i$ are variational parameters which can be used to minimize the energy.
An explicit numerical example is presented demonstrating that the mean-field energy using two orbitals can be lower
than the GP energy.
Since in the absence of interaction between the particles, all particles reside in a single orbital to form
the ground state of the condensate, it is clear that the present approach reproduces the GP result for weak interactions.
Beyond some critical interaction strength $\lambda=\lambda_0\,N$ obtained, for instance, 
by increasing the number $N$ of particles in the condensate, the mean-field energy can drop below the GP one.
This critical value depends strongly on the external potential. 
By varying the shape and parameters of this potential one can lower or raise the
interesting range of $\lambda$ substantially.

In our simple example it has been shown that the mean-field energy can drop below the GP energy 
for a continuous range of the occupation $n_i$.
The system then consists of two (or more) subsystems each possessing its own chemical potential. This may contradict
the picture one usually has of a condensate. It is, therefore, very interesting to note that these, in general different,
chemical potentials become identical at the optimal occupations $n_i$ which minimize the energy $E(\{n_i\})$ restoring
thereby the picture of a condensate.
This finding leads to another highlight of this work: self-consistent eigenvalue equations 
for the optimal orbitals at the optimal occupations could be derived.

It should be kept in mind that the occupation numbers $n_i$ are variational parameters in the present
mean-field and, therefore, the solution of eq.(10) will tell us via eq.(11) which the optimal occupation numbers
are. In those cases where one finds that $n_1/N=1$, then all bosons reside in a single orbital, our
ansatz reproduces the GP result and the GP mean field is thus the best mean field. 
The GP mean field is an excellent ansatz reflecting our physical
expectation that in the ground state all bosons would like to occupy the same orbital in order
to minimize the energy. 
It is, therefore, surprising that there exist counter examples at all.
Here, we have shown that an attractive condensate in an one-dimensional symmetric double-well potential provides
such a counter example. This finding is not restricted to the {\it specific shape} of the external 
symmetric double-well potential used here. Other counter examples can be expected. For instance, an attractive
condensate in an one-dimensional symmetric multi-well potential. Here, however, as many orbitals
must be included in the mean-field ansatz as there are wells.
That macroscopic condensation can be described by one-dimensional equations has been discussed in the
literature, see, e.g. \cite{13} and references therein.
Clearly, it would be very interesting to study condensates in higher dimensions.
Of course, the same equations as derived here hold also in this case, but their full numerical evaluation is
more involved and such a study is left to the future. 
We note that in
the present numerical example it is the symmetry of the problem which
fails the GP mean field. In other examples the reason that GP  is not the best mean field may be
of different origin. We suspect, however, that if one finds $n_1/N$ to deviate from 1, it is a helpful warning
that one might have to go beyond the mean-field approach as such in order to describe the underlying physics properly.

Being non-linear equations, eqs. (4) and (10) may have several solutions. We have encountered examples where the
lowest-energy solution is characterized by $n_1/N=1$, but other low-lying solutions exist with $n_1/N$ deviating from 1.
The first solution, of course, corresponds to the system's ground state. In the latter solutions the bosons reside in two
orbitals and their existence could be an indication for collective excited {\it self-consistent} states of the condensate.
We remind that according to eq. (8) there exists only a single value of the 
chemical potential for each of these excited states making them attractive to further investigations.

Finally, we would like to mention very briefly that (a) time-dependent mean-field equations corresponding to the
stationary equations discussed here are straightforwardly  derived, and (b) the present mean-field equations
provide an excellent starting point of investigating condensates beyond the mean-field approach.

Helpful discussions with Dieter Meyer and Ofir Alon are gratefully acknowledged.



\begin{figure}
\includegraphics[width=12.5cm]{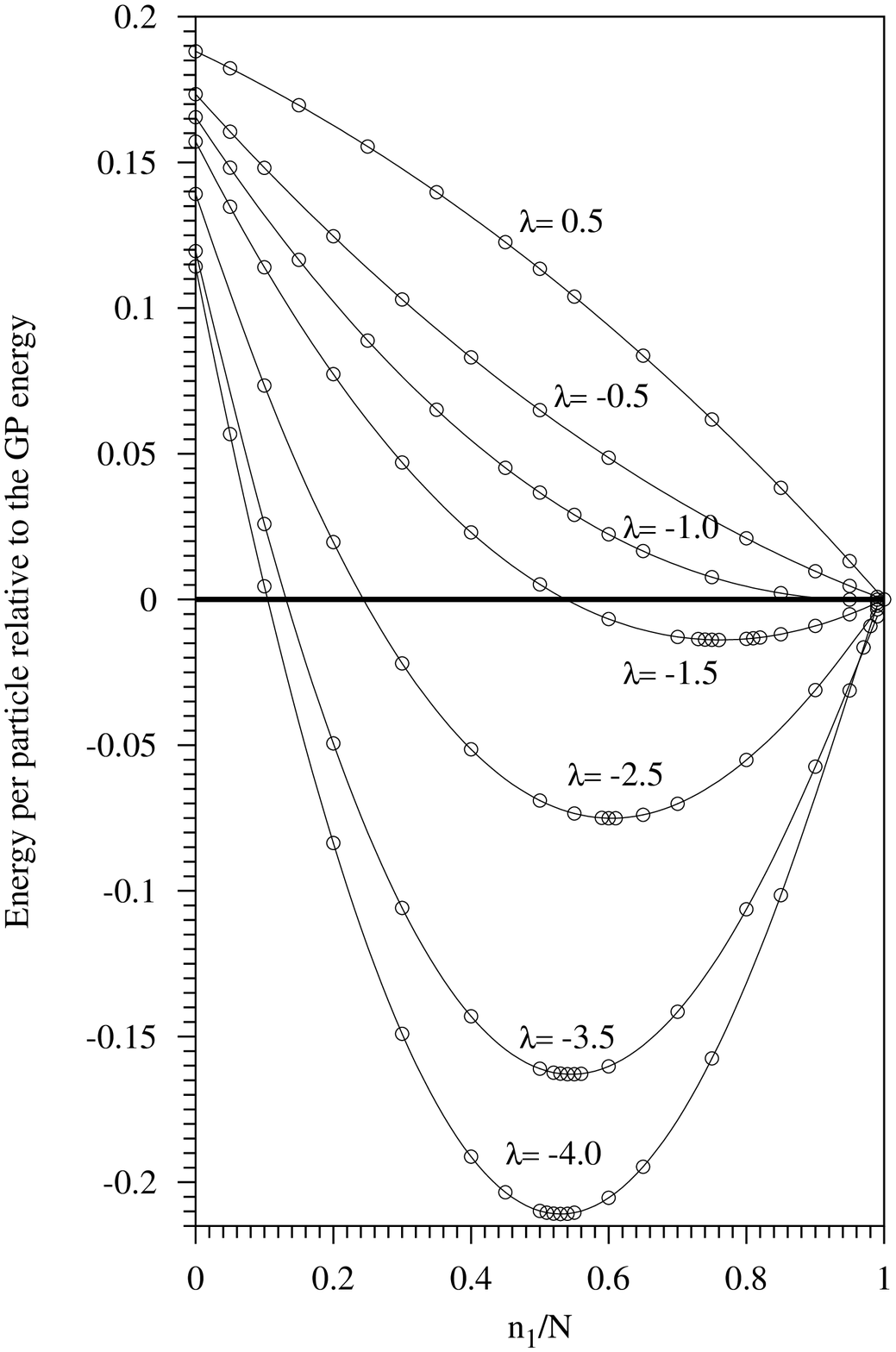}
\caption{
The difference between the present mean-field energy per particle $E(n_1)/N$ and the GP energy per particle $E_{GP}/N$
is shown for several values of $\lambda=\lambda_0\,N$ as a function of the fractional occupation $n_1/N$.
It is seen that the energy $E(n_1)$ may drop substantially below the $GP$ energy. At $n_1/N=1$ both energies,
of course, coincide. The external potential (4) has been used with $x_0=1.5$ and $a=1.5015$.
$\lambda$ and the energies are given in units of $\omega$}
\label{fig1}
\end{figure}

\begin{figure}
\includegraphics[width=12.5cm]{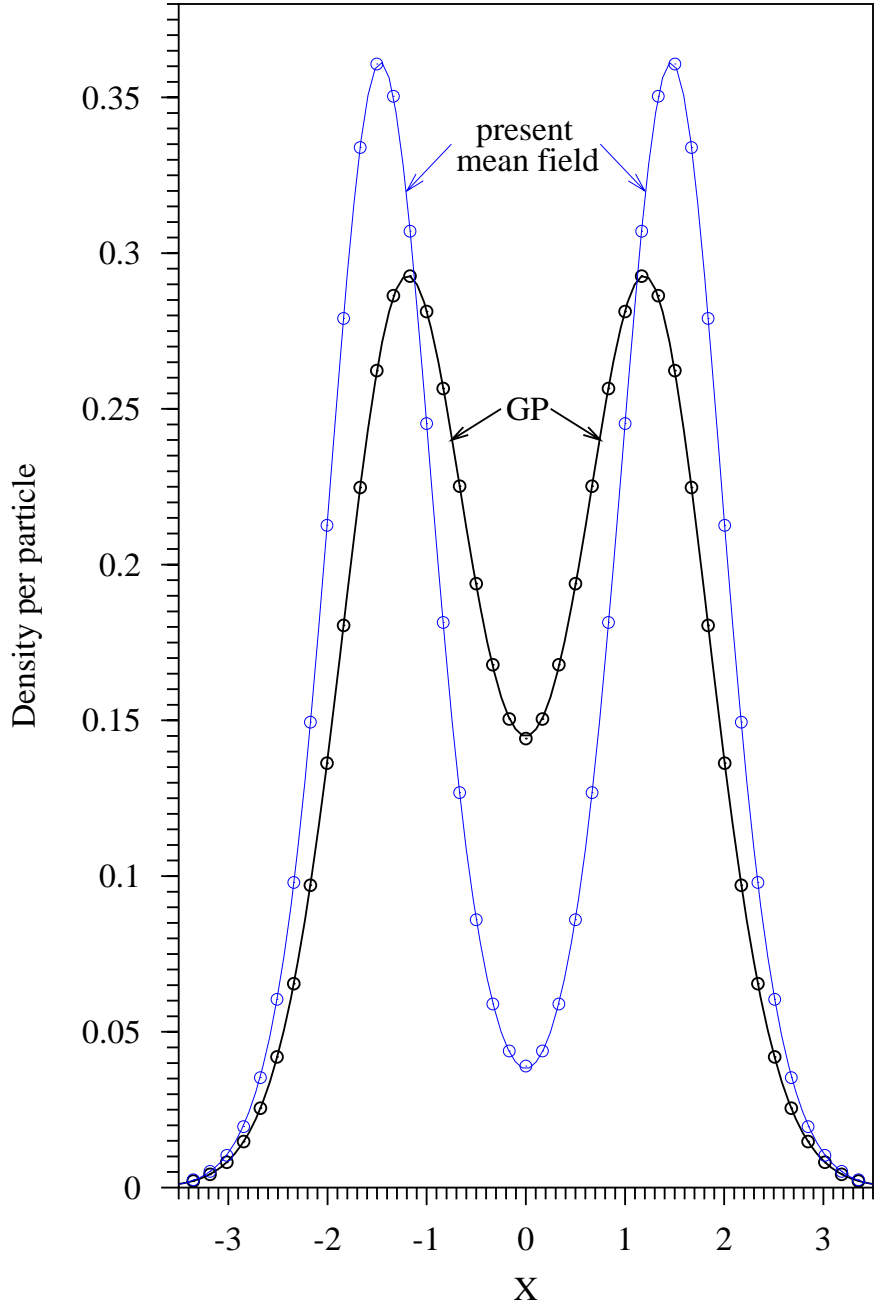}
\caption{
The density per particle $\rho=( n_1|\varphi_1|^2+ n_2|\varphi_2|^2)/N$
is shown for one of the examples of fig.1 ($\lambda=-2.5$ and $n_1/N=0.6$) and compared with
the corresponding $GP$ density per particle $\rho_{GP}= |\varphi|^2$.}
\label{fig2}
\end{figure}

\end{document}